\documentstyle[12pt,aaspp4,flushrt,epsf]{article}
\newcommand{\lya}{Ly$\alpha$}

\newcommand{\kms}{~{\rm km \ s}^{-1}}
\newcommand{\etal}{et al.\ }

\slugcomment{}

\lefthead{A. Ortiz-Gil \etal} 
\righthead{\lya\ absorbers in groups}

\begin{document}

\title{The gaseous extent of galaxies and the origin of \lya\
absorption systems. IV: \lya\ absorbers arising in a galaxy
group.\altaffilmark{1}}

\author{A. Ortiz-Gil\altaffilmark{2,3}, K.M. Lanzetta\altaffilmark{2}, 
J.K. Webb\altaffilmark{3}, X. Barcons\altaffilmark{4}, A. Fern\'andez-Soto
\altaffilmark{3}}

\altaffiltext{1}{Based on observations with the NASA/ESA Hubble Space
Telescope, obtained at the Space Telescope Science Institute, which is
operated by the Association of Universities for Research in Astronomy,
Inc., under NASA contract NAS5--26555.}

\altaffiltext{2}{Department of Physics \& Astronomy,
State University of New York at Stony Brook, Stony Brook, NY 11794--3800,
U.S.A.}

\altaffiltext{3}{School of Physics, University of New South Wales, 
Kensington, NSW-2031, Sydney, AUSTRALIA}

\altaffiltext{4}{Instituto de F{\'\i}sica de Cantabria (Consejo
Superior de Investigaciones Cient{\'\i}ficas--Universidad de
Cantabria), 39005 Santander, SPAIN}

\date{November 1998}

\clearpage

\begin{abstract}

We present new GHRS observations of \lya\ absorption lines associated
with a group of galaxies towards the QSO 1545$+$2101.  We have
identified eight distinct Ly$\alpha$ absorption features in the
spectrum of QSO 1545$+$2101 at a mean redshift of $<\!\!\! z \!\!\!> =
0.2648$ with a velocity dispersion of $163 \kms$.  A group of galaxies
is detected in the vicinity of this QSO at a mean redshift of $<\!\!\!
z \!\!\!> = 0.2645$ and velocity dispersion $239 \kms $.

The identification of discrete absorption systems indicates that they
arise in clouds of neutral hydrogen rather than in a diffuse
intragroup medium. Our analysis suggests that the \lya\ absorption
lines are associated with individual galaxies in the group, although
a one-to-one relationship between absorbers and galaxies is difficult
to establish in such a dense environment.

\end{abstract}

\clearpage

\section{Introduction}

\lya\ absorption due to groups or clusters of galaxies has only been
detected relatively recently (Lanzetta \etal 1996, Ortiz-Gil \etal
1997, Tripp \etal 1998, Shull \etal 1998).  This is because it had
been difficult to identify suitable combinations of background QSOs
and foreground clusters close enough to the QSO line of sight to
produce \lya\ absorption in the QSO spectrum. The search
for suitable groups was poorly motivated since \lya\ absorption would
not be expected due to the high temperature of the intracluster
medium. Furthermore the majority of Lyman-alpha absorption line data 
in the literature are at redshifts z$>1.5$ (when the line is shifted into 
the optical band), where most clusters are still on the process of 
formation.

Recent observations appear to produce contradictory results:
Morris \etal (1993) and Bowen, Blades, \& Pettini (1996) failed to
identify absorption due to clusters of galaxies, in contrast to the
results of Lanzetta, Webb \& Barcons (1996), Ortiz-Gil \etal (1997),
Tripp \etal (1998), and Shull \etal (1998).

However, other recent work suggests that galaxies may partially retain their
gaseous envelopes in a cluster environment (see Cayatte \etal 1990 for
an example in the Virgo cluster). Zabludoff \& Mulchaey (1998) find
that galaxies in poor groups ($\approx 20 - 50$ members) lie in a
common halo which contains most of the mass of the group.  Blitz \etal
(1998) suggest that the High Velocity Clouds (HVCs) detected in the
Local Group might be the counterparts of Lyman Limit systems, as they
find similar column densities and internal velocity dispersions and
subsolar metallicites.  They also suggest that lower column density
HVCs may correspond to \lya\ clouds.  In their models the HVCs trace
the distribution of dark matter in and around the group, following its
filamentary/sheet-like structure.
 
Lanzetta \etal (1996) report the identification of a group of galaxies
toward QSO 1545$+$2101 responsible for a broad absorption feature
present in an HST \sl Faint Object Spectrograph \rm (FOS) spectrum of
this object. The spectral resolution of these data was insufficient to
show whether it was the group as a whole or individual galaxies within
it which were responsible for the observed \lya\ absorption.
Individual galaxies would give rise to discrete absorption components
associated with particular galaxies. Also, galaxies at a smaller
impact parameter should produce higher column densities.

In this paper we present new higher resolution observations of the QSO
1545$+$2101 using the \sl Goddard High Resolution Spectrograph \rm on
HST.

Throughout this paper we have adopted a deceleration parameter
$q_0=0.5$ and a dimensionless Hubble constant $h=H_0/100 (\kms
\mbox{Mpc}^{-1})$.

\section{HST/GHRS observations of the QSO 1545$+$2101}
\label{data}

A high resolution spectrum of the QSO 1545$+$2101 was obtained using
the GHRS spectrograph with the G160M grating on August 23 1996
(Fig. \ref{fig1}). The observations were obtained during a series of
14 exposures each of 300s duration for a total exposure time of 4200s.
The individual exposures were reduced using standard pipeline
techniques and were registered to a common origin and combined using
our own reduction programs. The final spectrum was fitted with a
smooth continuum using an iterative spline fitting technique. The
spectral resolution of the final spectrum was measured to be
FWHM=$0.07$ \AA\ (or FWHM=$ 14 \kms$), and the continuum
signal-to-noise ratio was measured to be $S/N \approx 12$ per
resolution element. Previous imaging and spectroscopic observations of
the field surrounding QSO 1545$+$2101 are described in Lanzetta \etal
(1996).

\section{Detection and analysis of the absorption systems in the
field of QSO 1545$+$2101}

We detected all absorption lines above a significance level of $3
\sigma$ in the spectrum of QSO 1545$+$2101.  The parameters
characterizing each absorption line were estimated using
multi-component Voigt profile fitting (three different and
independently developed software packages were used, the three of them
providing the same results).  The values obtained are shown in Table
\ref{tab1}. The $3 \sigma$ confidence level detection limit was found
to be $0.03$~\AA\ .

Two absorption lines are found at $z=0.2504707 \pm 0.0000030$ and
$z=0.2522505 \pm 0.0000016$, with a velocity separation of $\sim 427\,
\kms$. A group of at least eight lines is also detected, the redshift
centroid of this group being $<\!\!\! z \!\!\!>= 0.2648 \pm 0.0002$ with a velocity
dispersion of $163\pm 57 \kms$.

Galactic heavy element lines in the spectrum (C {\sc iv}
$\lambda\lambda$1548,1550 and Si {\sc ii}$\lambda$ 1526) were
identified and removed them from subsequent analyses (Fig.\ref{fig1}).

\section{Group of galaxies in the field of QSO 1545+2101}

The spectroscopic galaxy sample was selected on the basis of galaxy
brightness and proximity to the QSO line-of-sight.  Although this
sample cannot be considered complete the galaxies were selected
randomly from an essentially flux-limited sample. The sample of
galaxies used in this study is given in Table \ref{tab2}.  Figure
\ref{fig0a} is an image of the field.  The galaxies in the group which
were observed spectroscopically are indicated in that figure. The
error in the galaxies redshift is estimated to be $ \sim 120\, \kms$
(Lanzetta \etal 1995).

We detected a group of seven galaxies with redshift centroid at
$<\!\!\! z \!\!\!>=0.2645 \pm 0.0004$ and velocity dispersion of
$239\pm 90 \kms $. Their individual impact parameters range from $7.2$
up to $456.4 \ h^{-1}$ kpc. 

We searched in the archives for X-ray data on this field as
detecting X-ray emission from the group  might help to
characterize it better.  Unfortunately the group is so close to
the QSO (which is an X-ray source itself) that only the {\it ROSAT}
High Resolution Imager data would be of any use, and there are no such
data in the archive. QSO 1545$+$2101 was observed both with the {\it
Einstein} Imaging Proportional Counter and with the {\it ROSAT}
Position Sensitive Proportional Counter, but the extended
Point-Spread-Function of both instruments resulted in QSO emission
severely contaminating the region where the galaxy group might
emit X-rays.

In any case, from our image of this field and the fact that the
velocity dispersion that we have measured for this group is quite
small, it is clear that it is a loose association of galaxies rather
than a galaxy cluster. The maximum value of impact parameter in our
sample ($456.4 \ h^{-1}$ kpc) is also the typical physical size for
poor groups of galaxies (Zabludoff \& Mulchaey 1998).  In fact, this
group is most probably the one hosting the QSO itself, as
$z_{em}=0.264 \pm 0.0003$ for this object (Marziani \etal 1996).  One
might therefore expect this group to have only between 10-20 members
(Bahcall \etal 1997).

\section{The relationship between the absorption systems and the galaxies}
\label{relation}

The cluster of absorption lines present in the spectrum of QSO
1545+2101 might arise in different environments. They might be intrinsic
absorbers, arising either in the QSO region itself or in its near
environment.  The similar redshifts of the QSO and the group of
absorbers in these data may point to one of these hypotheses as the
right one.  In addition, QSO 1545+2101 is a radio-loud object and
there have been suggestions that intrinsic absorption or absorption
arising in the QSO host galaxy would be stronger in radio-loud QSOs
than in radio-quiet ones (Foltz \etal 1988; Mathur, Wilkes \& Aldcroft 1997).

But there are also some arguments against an intrinsic origin for
these lines, as discussed by Lanzetta \etal
(1996). Associated absorbers of the kind described above produce
rather strong absorption lines, while the lines detected in QSO
1545+2101 are relatively weak. Moreover, no corresponding metal
absorption lines have been detected. With the
new data from GHRS we have more evidence against the
associated nature of the absorption systems: the agreement between
the galaxies' and absorbers' redshift centroids and between their
corresponding velocity dispersions suggests that both galaxies and
absorbers groups share---at least---the same physical location.  
These characteristics lead us to reject the associated hypothesis
and to consider another possible scenario:
absorption arising in cosmologically intervening objects, 
within the same group of galaxies that hosts the QSO.

To further explore this third hypothesis a demonstration of
the non-random coincidence between the galaxies and absorbers positions
in velocity space is necessary. Identifying each galaxy with a single 
absorption line would also be very interesting. In what follows we
use statistical methods to address both questions.

\subsection{A cluster of absorbers arising in a group of galaxies}
\label{random}

The group of galaxies detected toward the QSO 1545$+$2101
has a mean redshift of $<z_g>=0.2645 \pm 0.0004$, compatible with the
mean redshift centroid value of the group of absorption lines,
$<z_a>=0.2648 \pm 0.0002$ (to the red of the Galactic Si {\sc ii} line
in Fig. \ref{fig1}).  The absorber and galaxy velocity dispersions are
similar: $239\pm 90 \kms$ for the group of galaxies and $163\pm 57
\kms$ for the group of absorption lines (the error in the galaxy 
redshifts is $\sim 120 \kms$ ). This strongly suggests a
connection between the absorbers and galaxies.

There is also a galaxy in this field whose redshift is $z=0.2510$.
Two \lya\ absorption lines are detected near this redshift at
$z=0.2504707 \pm 0.0000030$ and $z=0.2522505\pm 0.0000020$ (to the
blue of the Galactic Si {\sc ii} line in Fig. \ref{fig1}). The
galaxy-absorber velocity differences are $ \Delta v = 127\pm 120 \kms$
and $\Delta v = 300\pm 120\, \kms$ respectively. This implies that
this galaxy, whose impact parameter is $\rho=306.4 \ h^{-1}$ kpc,
could be responsible for one of the absorption lines as both $\Delta
v$ values are compatible with the velocity dispersions that one finds
typically in a galactic halo ($\approx 200 \kms$). The velocity
difference between the two absorption systems ($427 \kms$), is perhaps
too large for the same galaxy to be responsible for both of them. It
may also be that we have not observed the actual galaxy giving rise to
either of these absorption lines, since our galaxy sample is not
complete.

\subsubsection{Statistical analysis}

Two statistical tests were carried out to investigate the relationship
between the group of galaxies and absorbers.  In the first, we
computed the two-point cross-correlation function ($\xi_{ag}$) between
the absorbers and the galaxies.  This function was normalized by
computing the $\xi_{ag}$ expected if there were no relation between
absorbers and galaxies (derived using galaxy redshifts which are
randomly distributed over a redshift range around the real group of
absorption lines).  Errors were computed using a bootstrap method,
simulating 1000 samples of 7 galaxy redshifts, each set randomly
selected from the real set, deriving error estimates from the
distribution of 1000 values of $\xi_{ag}$. The final result is
shown in Fig. \ref{fig3} (panel $a$).

In the second statistical test, we applied a similar statistical
method, not to the actual set of absorbers but to the individual pixel
intensities in the spectrum.  This way we overcome any potential
problems introduced as a consequence of incorrect determination of the
true velocity structure in the profile fitting process, or the
presence of weak lines falling below the detection threshold.  Recent
analyses by Liske, Webb \& Carswell (1999) show that the study of
pixel intensities is more sensitive to clustering than the usual
line--fitting techniques.  The test is as follows: we evaluated for
each pixel $i$ the value of the function $g_i=1-f_i$, where $f_i$ is
the intensity of pixel $i$ normalized to the continuum.  Clearly,
$g_i$ has larger values in pixels belonging to absorption lines.  For
pixels corresponding to metal lines (including galactic lines) we
assigned a value of $g_i=0$. Then, for each galaxy we compute the
function $\zeta_v= \sum_{n=1}^{N} \sum_j g_{nj}$, where $g_{nj}$
refers to all the pixels $j$ located at a distance $v$ in velocity
space from galaxy $n$, $N$ being the total number of galaxies in the
sample. The velocity distances $v$ considered range from $0 \kms$ up
to the one spanned by the whole spectrum. This function $\zeta_v$ is
analogous in some sense to the two-point cross-correlation function
computed before: high values of $\zeta_v$ at low velocity distances
reflect the tendency of the pixels corresponding to absorption lines
to lie close to the galaxies. The same function $\zeta_v$
corresponding, again, to a randomly selected sample of galaxies chosen
from a uniformly distributed group was computed in a way analogous to
the previous one and the result, after normalizing to this random
case, is shown in Fig. \ref{fig3} (panel $b$). The error bars
in $\zeta_v$ were computed using a bootstrap method as before.  The
result obtained is similar to the one obtained by
computing the two-point cross-correlation function: most of the pixels
belonging to absorption lines (i.e., with larger values of $g$) lie
less than $200 \kms$ away from the galaxies.

\subsection{Are absorption lines related to galaxies on a case-by-case basis?}
 
As there is a clear complex of discrete \lya\ lines in the spectrum of
QSO 1545+2101, we explored the possibility of a one-to-one match
between galaxies and absorbers.

A Gaussian model was assumed for the distribution of galaxies in the
group. The null hypothesis is that the galaxies are randomly drawn
from a Gaussian whose parameters are derived from the real data.  The
two-point correlation functions $\xi_{ag}$ and the function $\zeta_v$
were computed in the same way as in \S \ref{random} (see Fig.
\ref{fig3}, panels $c$ and $d$).  No evidence of a one-to-one
correspondence between galaxies and absorbers is found.

We can estimate the maximum statistical velocity dispersion between
individual absorbers and their galaxies of origin that would permit
the detection of a one--to--one correspondence, assuming an intrinsic
one--to--one correspondence exists.  A Kolmogorov-Smirnov Monte Carlo
test showed that if the average velocity dispersion between the
galaxies and the corresponding absorbers were $\lesssim 100 \kms$ then
the null hypothesis would be rejected ($>2 \sigma$) in $90$\% of the
cases.  As a typical galactic velocity dispersion
is  $\sim 200 \kms$, this condition is not likely to be satisfied in
practice. Another possibility is to improve the statistics. From Monte Carlo
simulations we estimate that about 100 groups similar to the ones
studied here are needed for a $2 \sigma$ detection of this one--to--one
association, for a velocity dispersion between the absorber and the
galaxy of about $200 \kms$.

\section{Discussion and conclusions}
\label{discussion}

We have detected a clump of absorption lines along the line-of-sight
towards the QSO 1545$+$2101.  A group of galaxies has also been
detected, with impact parameters $\rho$ of the individual galaxies to
the QSO line--of--sight of less than $\sim 460 h^{-1}$ kpc. The group
is probably the one hosting the QSO, so we can expect it to have about
10-20 members (Bahchall \etal 1997).

Several scenarios might give rise to the absorption.  Due to the close
redshift values of the QSO and the group of absorbers one could easily
think that they arise either in the QSO itself or in the corresponding
host galaxy.  We consider that there are compelling arguments
supporting the intervening system hypothesis (see Lanzetta \etal 1996)
and contradicting the associated hypothesis. We now summarise those
arguments.

The velocity spanned by the group of \lya\ absorption lines is
consistent with the velocity dispersion of the group of galaxies. This
implies that the \lya\ absorbers arising in that group occupy the same
region of space as the galaxies themselves.  Moreover, the spectrum of
QSO 1545$+$2101 reveals a group of \it discrete \rm \lya\ absorption lines
at $z \approx 0.26$.  Multi-component Voigt profile fitting provides a
statistically good fit to the data, indicating that the absorption
lines arise in overdense gas regions rather than in some smoothly
distributed intragroup medium.

The average Doppler dispersion parameter of the absorption lines, {\it
b}, is measured to be $19\pm 4 \kms$ with a dispersion of $10\pm 4
\kms$. This value is in agreement with the values measured in the low
redshift \lya\ forest. Therefore there is no evidence from this case
for any physical difference, in terms of the {\it b}
parameter, between \lya\ clouds lying within or outside of groups.

Two statistical analyses show that the distribution of galaxies with
respect to the absorbers is not random, but that is not possible to
confirm a one--to--one match due to the proximity in velocity space of
the galaxies in the groups and the uncertainty on their redshifts. A
Kolmogorov-Smirnov test showed that a small galaxy-absorber velocity
dispersion (less than $\sim 100 \kms$) would be required to establish
a one--to--one match. As this is well below the typical values
corresponding to a galaxy potential well, another approach is
required, such as having a large enough sample of clusters or groups
of galaxies related to clusters of \lya\ absorption lines.  All the
facts above support the idea of a physical connection between the
group of galaxies and the group of \lya\ absorbers.

Another piece of circumstantial evidence pointing to a one--to--one
relationship between absorbers and galaxies.  This concerns the number
of each type of object detected.  As mentioned before, the galaxy
group towards QSO 1545$+$2101 contains approximately 10-20 members.
According to Lanzetta \etal (1995), only a subset of them will be
close enough to the QSO line of sight to produce observable \lya\
absorption ($\rho \lesssim 160$ kpc for a covering factor of $\approx
1$).  We observe eight individual components in the absorption
profile, consistent with the expectations from such a na\"{\i}ve
model. There is no obvious reason why such an agreement would be found
for some other quite different model (be it HVCs, filaments or any
other structure).  We note that the absorption lines may break up into
further components at higher spectral resolution although these may
then be substructure within individual galaxies.

\vskip 1truecm

\acknowledgments

A.O.-G., K.M.L. and A.F.-S. were supported by grant NAG-53261; grants
AR-0580-30194A, GO-0594-80194, GO-0594-90194A and GO-0661-20195A from
STScI; and grant AST-9624216 from NSF. A.O.-G. acknowledges support from
a UNSW Honorary Fellowship. A.F.-S. was also supported by
an ARC grant.  X.B. was partially supported by the DGES under project
PB95-0122.

\newpage
\setcounter{figure}{0}

\figcaption[espgal2_1545.ps]{Spectrum of QSO 1545$+$2101, smoothed at
the Nyquist rate. The original data have a resolution of FWHM$=0.07
\AA$ (FWHM$=14 \kms$). Tick marks in the upper panel indicate the
predicted wavelengths of \lya\ at the redshifts of the galaxies listed
in Table \ref{tab2}. Tick marks in the lower panel show the positions
of the detected absorption lines}

\figcaption[image1545_mark_neg.eps] { Image of a $\approx 12' \times
12'$ field toward QSO 1545$+$2101 (north is up, east is left). Marked
galaxies are the ones belonging to the group. Numbers increase with
increasing z. The QSO is marked by number $0$}

\figcaption[correl.ps.] {(a): cross-correlation function between
galaxies and absorbers
normalized using a uniform
distribution. The existence of a correlation between them
is clear. (b): normalized $\zeta_v$
function (see text).
There is a clear connection between the
real sample of galaxies and the pixels corresponding to absorption
lines. (c): cross-correlation function between galaxies and
absorbers 
normalized using a Gaussian distribution. No significant correlation
between them is found. (d):
normalized $\zeta_v$ function (see text). A one-to-one match
between the absorption systems and the galaxies cannot be established}

\clearpage

\begin{table}[t]
\begin{center}
\begin{tabular}{c c c c c c c c}
\tableline
$z_{abs}$ & $\sigma_z$ & $W$ & $\sigma(W)$& $b$& $\sigma(b) $ & $\log N$  & $\sigma(\log N)$ \\ 
 & ($10^{-7}$) & (\AA) & ($10^{-4}$\ \AA) & $(\kms)$ & $(\kms)$ &($\mbox{cm}^{-2}$) & ($\mbox{cm}^{-2}$) \\
\tableline
0.2504707 & 30 & 0.067 & 0.013 & 66    & 9   & 13.61 & 0.05 \\
0.2522505 & 16 & 0.21  & 0.03  & 40    & 5   & 13.63 & 0.04 \\
0.2634336 &  3 & 0.387 & 0.018 & 28.8  & 1.4 & 14.17 & 0.04 \\
0.2641748 &  3 & 0.307 & 0.016 & 28.0  & 1.2 & 13.94 & 0.02 \\
0.2645921 &  4 & 0.294 & 0.015 & 27.3  & 1.3 & 13.91 & 0.02 \\
0.2648250 &  6 & 0.033 & 0.009 &  4    & 5   & 12.84 & 0.12 \\ 
0.2650173 & 10 & 0.254 & 0.015 & 26    & 3   & 13.81 & 0.05 \\ 
0.2652287 & 23 & 0.117 & 0.014 & 23    & 6   & 13.33 & 0.12 \\
0.2654554 &  9 & 0.041 & 0.012 &  11   & 4   & 12.84 & 0.08 \\ 
0.2656921 &  4 & 0.203 & 0.017 & 23.4  & 1.5 & 13.67 & 0.03 \\
\tableline
\end{tabular}
\end{center}
\caption{\lya\ absorption systems toward QSO 1545$+$2101. 
For each absorption line,
$z_{abs}$ is the redshift, $W$ is the rest-frame equivalent width, $b$ 
is the Doppler parameter and $N$ is the column density}
\label{tab1}
\end{table}

\clearpage

\begin{table}[b]
\begin{center}
\begin{tabular}{c c c c c c c}
\tableline
$\Delta \alpha$&$\Delta \delta$&$\theta$&$ R$ &$z_{gal}$&$\rho$&$M_{R}-5\log h$\\
(arcsec) & (arcsec) & (arcsec) & & & ($h^{-1}$\ kpc)& \\
\tableline
-38.7 & -118.3 & 124.5 & 20.1 & 0.2510 & 306.4 & -19.4 \\
58.3  &   31.2 &  66.1 & 19.2 & 0.2630 & 167.7 & -20.4 \\
53.4  & -171.5 & 179.6 & 20.2 & 0.2638 & 456.4 & -19.4  \\
 7.0  &  166.5 & 166.6 & 20.6 & 0.2639 & 423.5 & -19.0 \\
16.6  &   -8.7 & 18.7  & 19.9 & 0.2639 & 47.6  & -19.7  \\
85.2  &  -81.3 & 117.8 & 18.8 & 0.2652 & 300.2 & -20.8  \\
-2.7  &   -1.1 &   2.9 & 19.1 & 0.2657 &   7.2 & -20.5  \\
-21.8 &   98.8 & 101.2 & 18.8 & 0.2658 & 258.3 & -20.8 \\
\tableline
\end{tabular}
\end{center}
\caption{Group of galaxies in the field of QSO 1545$+$2101. For each galaxy,
 $\Delta \alpha$ and $\Delta \delta$ are the offset from the QSO in right 
ascension and declination, $\theta$ is the angular offset from the
QSO, $R$ is the observed $R$-band magnitude, $z_{gal}$ 
is the redshift, $\rho$ is the impact parameter to the QSO line of sight and 
$M_R$ is the absolute $R$-band magnitude}
\label{tab2}
\end{table}

\clearpage
\setcounter{figure}{0}

\begin{figure}
\def\epsfsize#1#2{1.\hsize}
\centerline{\epsffile{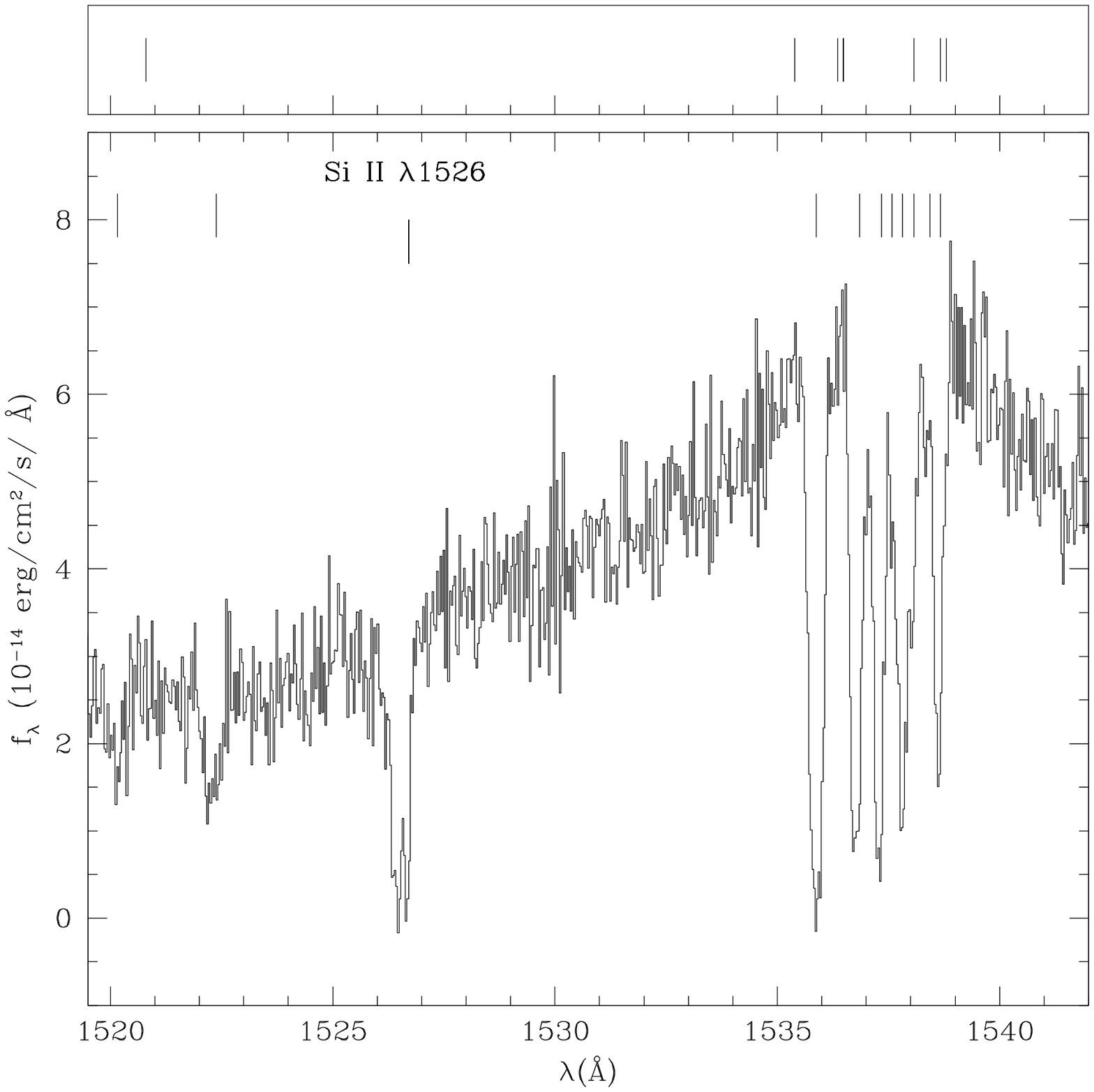}}
\caption{}
\centerline{\label{fig1}}
\end{figure}

\clearpage

\begin{figure}
\def\epsfsize#1#2{1.\hsize}
\centerline{\epsffile{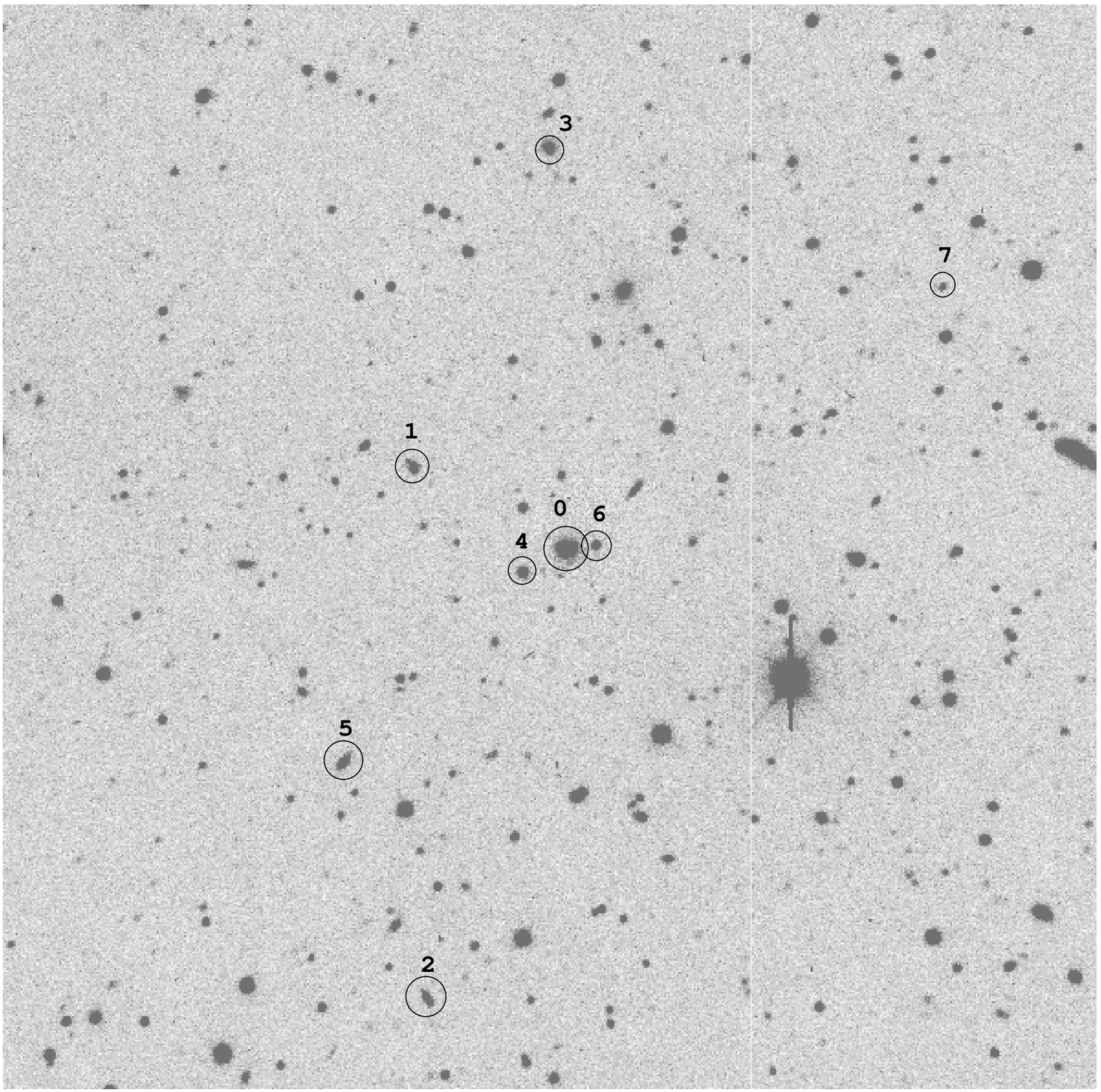}}
\caption{}
\centerline{\label{fig0a}}
\end{figure}

\clearpage

\begin{figure}
\def\epsfsize#1#2{1.\hsize}
\centerline{\epsffile{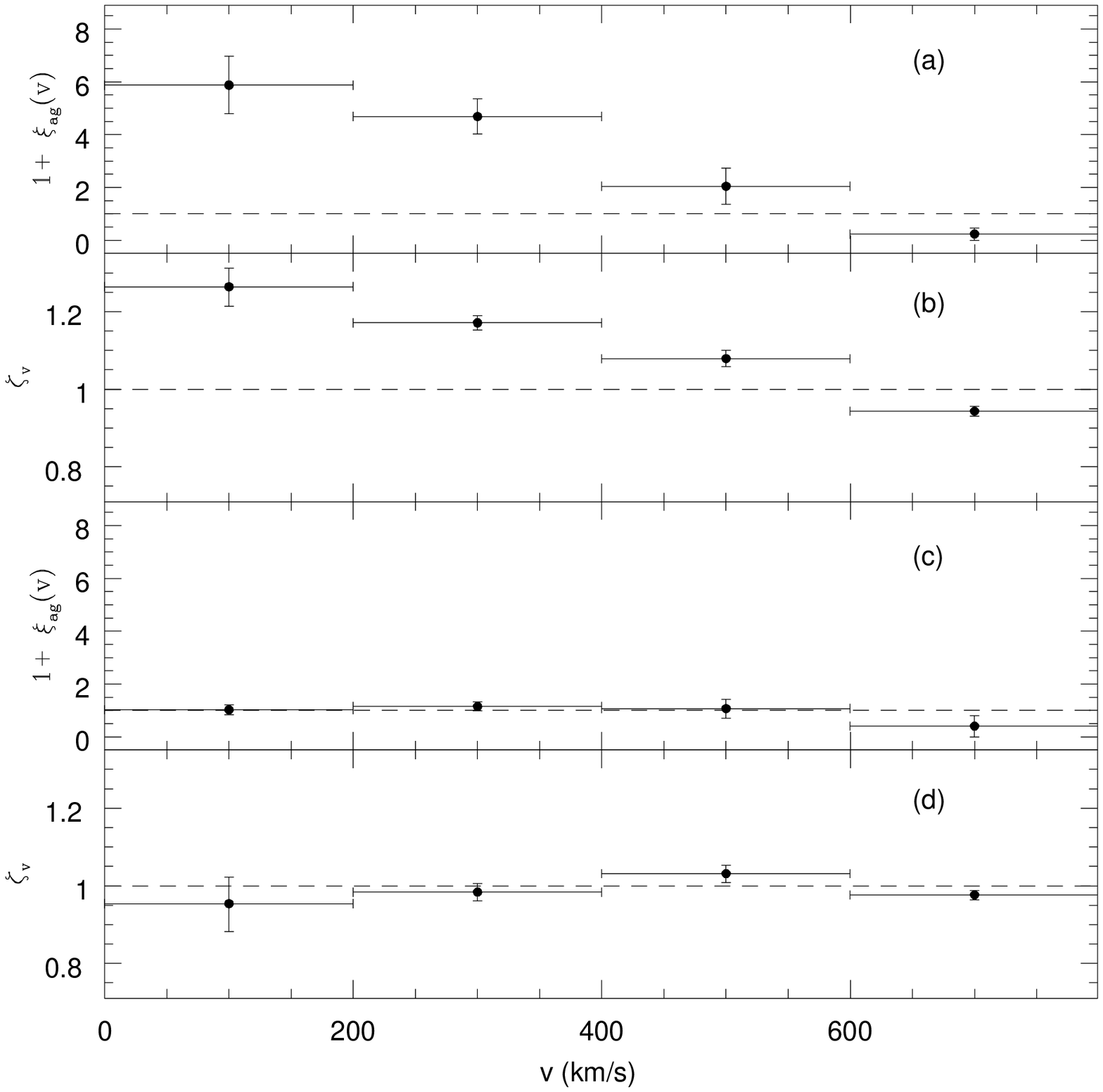}}
\caption{}

\centerline{\label{fig3}}
\end{figure}

\end{document}